\documentclass[aip, apl, reprint, longbibliography, nofootinbib]{revtex4-1}
\usepackage{xr-hyper, hyperref, graphicx, amsmath, amssymb, xcolor}

\begin{document}

\title{Multifunctional 2d infrared photodetectors enabled by asymmetric singular metasurfaces}

\author{Valentin Semkin}
\email[]{semkin.va@phystech.edu}
\affiliation{Moscow Institute of Physics and Technology, Dolgoprudny 141700, Russia}

\author{Aleksandr Shabanov}
\affiliation{Moscow Institute of Physics and Technology, Dolgoprudny 141700, Russia}
\affiliation{Joint Stock Company ''Skanda Rus'', Krasnogorsk 143403, Russia}

\author{Kirill Kapralov}
\affiliation{Moscow Institute of Physics and Technology, Dolgoprudny 141700, Russia}
\affiliation{Joint Stock Company ''Skanda Rus'', Krasnogorsk 143403, Russia}

\author{Mikhail Kashchenko}
\affiliation{Moscow Institute of Physics and Technology, Dolgoprudny 141700, Russia}
\affiliation{Programmable Functional Materials Lab, Center for Neurophysics and Neuromorphic Technologies, Moscow 127495, Russia}
\affiliation{Joint Stock Company ''Skanda Rus'', Krasnogorsk 143403, Russia}

\author{Alexander Sobolev}
\affiliation{Moscow Institute of Physics and Technology, Dolgoprudny 141700, Russia}
\affiliation{Joint Stock Company ''Skanda Rus'', Krasnogorsk 143403, Russia}

\author{Ilya Mazurenko}
\affiliation{Moscow Institute of Physics and Technology, Dolgoprudny 141700, Russia}

\author{Vladislav Myltsev}
\affiliation{Moscow Institute of Physics and Technology, Dolgoprudny 141700, Russia}

\author{Egor Nikulin}
\affiliation{Russian Quantum Center, Skolkovo, 143025 Moscow, Russia}

\author{Alexander Chernov}
\affiliation{Moscow Institute of Physics and Technology, Dolgoprudny 141700, Russia}
\affiliation{Russian Quantum Center, Skolkovo, 143025 Moscow, Russia}

\author{Ekaterina Kameneva}
\affiliation{Moscow Institute of Physics and Technology, Dolgoprudny 141700, Russia}

\author{Alexey Bocharov}
\affiliation{Joint Stock Company ''Skanda Rus'', Krasnogorsk 143403, Russia}

\author{Dmitry Svintsov}
\email[]{svintcov.da@mipt.ru}
\affiliation{Moscow Institute of Physics and Technology, Dolgoprudny 141700, Russia}
\affiliation{Joint Stock Company ''Skanda Rus'', Krasnogorsk 143403, Russia}

\begin{abstract}
Two-dimensional materials offering ultrafast photoresponse suffer from low intrinsic absorbance, especially in the mid-infrared wavelength range. Challenges in 2d material doping further complicate the creation of light-sensitive $p-n$ junctions. Here, we experimentally demonstrate a graphene-based infrared detector with simultaneously enhanced absorption and strong structural asymmetry enabling zero-bias photocurrent. A key element for those properties is an asymmetric singular metasurface (ASMS) atop graphene with keen metal wedges providing singular enhancement of local absorbance. The ASMS geometry predefines extra device functionalities. The structures with connected metallic wedges demonstrate polarization ratios up to 200 in a broad range of carrier densities at a wavelength of 8.6 $\mu$m. The structures with isolated wedges display gate-controlled switching between polarization-discerning and polarization-stable photoresponse, a highly desirable yet scarce property for polarized imaging.
\end{abstract}
\maketitle

High responsivity and low noise are necessary for electromagnetic detectors independent of the wavelength range. Fast response is highly desirable as well, especially in the optical communication realm. The latter is typically ensured either by ultimate detector scaling, or by exploiting novel high-mobility low-dimensional materials~\cite{yoshioka_ultrafast_intrinsic_detection,Massicotte_ps_response,Mueller2010,Muravev_response_time}. The needs for multiplexed telecommunications~\cite{ran_integrated_2021} and in-depth analysis of optical scenes~\cite{lu2020_hyperspectral,yoon2022hyperspectral} drive further demand for spectrum- and polarization-resolving photodetectors (PDs). Such resolution currently appears at the level of individual PDs, without sophisticated dispersion and polarization optics elements~\cite{Cai_tunable_spectrometers}. This becomes possible with the advent of computational optoelectronics~\cite{Geometric_deep_optical_sensing}. In this paradigm, the spectral and polarization sensitivity of individual PDs are tuned either with composition~\cite{Cai_NW_spectrometers}, or with geometry~\cite{Jiang_Multidimensional_PDs,Wei_Calibration_free,Wei_Configurable,Wei_Spin_light}, or, most preferably, with electrical biasing~\cite{Cai-spectrometer-1,Cai-spectrometer-2,Muravev_Interferometer,Black-P-spectrometer,Intelligent_IR_TBG,Kong_supercond_spectrometer}. The multi-dimensional arrays of photoresponse data are digitally processed to resolve the necessary characteristics of light.

Most reported spectrally- and polarization-tunable detectors represent piece-goods. Their material platforms, such as twisted bialyers~\cite{Intelligent_IR_TBG}, van der Waals heterostructures~\cite{Cai-spectrometer-1,Cai-spectrometer-2}, and encapsulated 2d flakes~\cite{Black-P-spectrometer} are far from scalable production. Solutions based on traditional metals~\cite{Kong_supercond_spectrometer} and III-V heterostructures~\cite{Muravev_Interferometer} required cryogenic temperatures. Achieving tunable photodetection with large-scale commercially available materials still represents a challenge for nanotechnology.

\begin{figure}[ht]
    \includegraphics[width=1\linewidth]{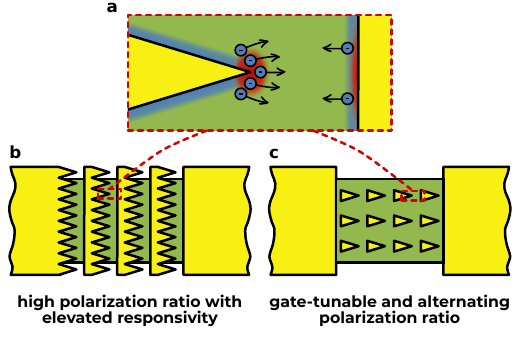}
    \caption{{\bf Asymmetric singular metasurface detectors.} {\bf a,} Each cell of the detector exploits singular local field enhancement at a wedge-shaped metal-2d contact, leading to an intense generation of photocarriers. The partial photocurrents at wedged and linear junctions are non-compensated, resulting in zero-bias response. {\bf b,c,} Individual cells are merged into metasurfaces for signal combining. Depending on merging strategy, the devices provide either very high or gate-tunable polarization ratios}
    \label{fig1}
\end{figure}

Here, we realize a methodology for zero-bias infrared detectors with configurable polarization response based on large-scale 2d materials with specially designed metallic patterns. A key property of these patterns is the presence of keen wedge providing singularly-enhanced highly non-uniform local optical field~\cite{Meixner1972,Pendry_Singular_MS}\footnote{Despite similar terminology, our structure is different from singular plasmonic graphene metasurfaces formed by smooth periodic doping, Ref.~\onlinecite{Galiffi_singular_MS_periodic_doping}}. The field non-uniformity drives the zero-bias photocurrent, as shown in Fig.~\ref{fig1} (a), while local field enhancement elevates the responsivity~\cite{detector_patent,wedge_zhang_2024}. Further enhancement is ensured by merging the individual wedges into {\it asymmetric singular metasurfaces} (ASMS), as shown in Fig.~\ref{fig1} (b). The magnitude of local field depends both on light polarization and wedge angle, which ensures geometric tuning of polarization ratio (PR) up to $\sim 200$. More surprisingly, we observe the gate tuning of PR between large positive and large negative values for other ASMS geometries with unconnected wedges [Fig.~\ref{fig1} (c)] and slit-patterned contacts. The ASMS-detectors belong to the class of geometrically-designed devices~\cite{Semkin_NL}, where the zero-bias response is ensured by contact geometry rather than $p-n$ doping~\cite{Doping_PNJ} or dissimilar metallization~\cite{Cai_Sensitive_THz,Mueller2010}.

\begin{figure}[ht]
    \includegraphics{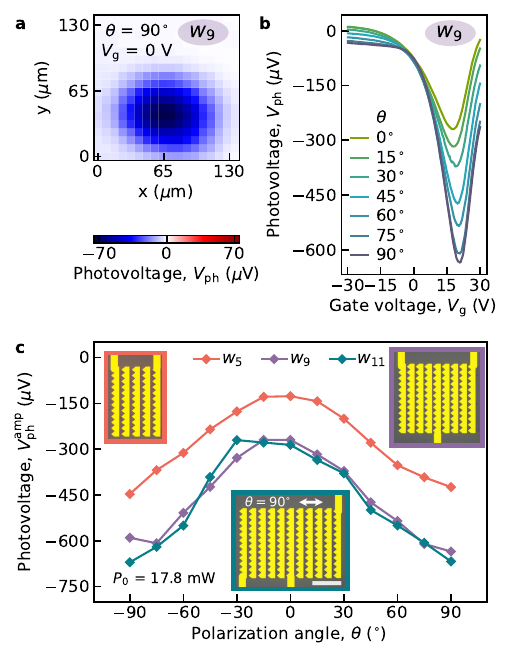}
    \caption{{\bf Proof of principle for ASMS detectors.} {\bf a,} Photovoltage map of the device $w_9$ proving stable zero-bias response. {\bf b,} Gate- and polarization dependencies of the photovoltage demonstrating a strong maximum at $\theta = 90^{\circ}$, i.e. for ${\bf E}$-field along the wedges. {\bf c,} Comparison of photovoltages between five-, nine- and eleven-stage detectors, shown with orange, violet and blue lines, respectively. Device micro-photographs are shown in the insets.  Scale bar is 20~$\mu$m.}
    \label{fig2}
\end{figure}

{\bf Principle of asymmetric metasurface detector.} We start our consideration with ASMS photodetectors shown in Fig.~\ref{fig2}. They comprise a CVD-graphene channel wet-transferred on Si/285 nm SiO$_2$ substrates, and an Au wedged metasurface directly atop the graphene. The first series contains three devices with five, nine and eleven cells in sequence, accordingly labeled as $w_5$, $w_9$ and $w_{11}$. The width of MS is 46 $\mu$m for all devices, and the length varies from 30 to 65 $\mu$m depending on the number of cells (other geometrical parameters listed in Fig.~\ref{resp_table}). The leftmost and rightmost MS contacts are used for electrical measurements of dc resistance and photovoltage. Prior to the measurements, the devices are evacuated to $10^{-4}$ Torr and electrically annealed.

The photovoltage signals are measured upon illumination with mid-infrared ($\lambda_0 = 8.6$ $\mu$m) quantum cascade laser radiation. The radiation is electrically modulated with frequency $f_{\rm las} = 911$ Hz, and the photovoltage signal is locked-in to this modulation. We allow for power-preserving polarization control by passing the laser beam through a quarter-plate and a polarizer. The radiation is focused into the spot with diameter $2\sigma \approx 40$ $\mu$m, and the total peak power in the spot $P_0$ is varied between 3.04~mW and 17.8 mW depending on the laser current.

Figure \ref{fig2}a shows the photovoltage map of the device $w_9$ and proves the stable detector response under uniform illumination. The photovoltage has the same sign, independent of the beam position $(x,y)$. It implies that the signal emerges due to the well-pronounced device asymmetry. This behavior persists at all carrier densities controlled by the gate, except for those where the signal is very weak.

\begin{figure*}[ht]
    \includegraphics{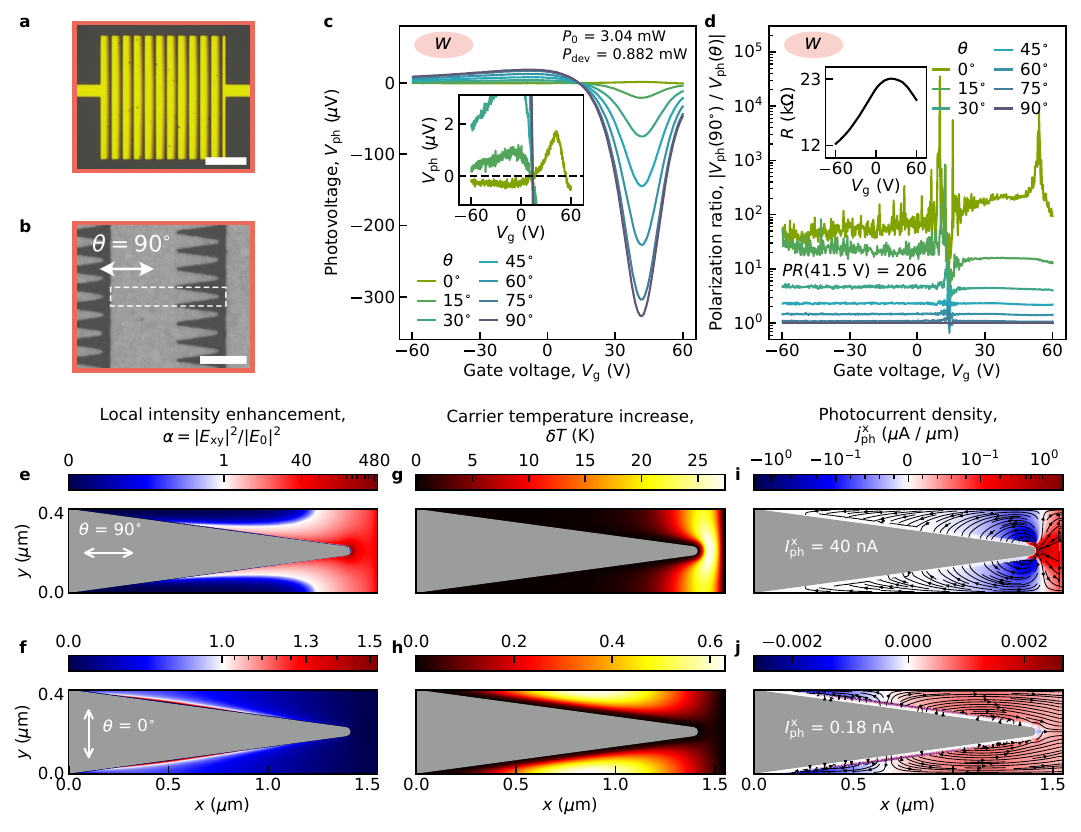}
    \caption{{\bf Optimized ASMS photodetector.} {\bf a,b,} Optical and scanning electron microscope images of fabricated structure. Scale bars are 10~$\mu$m and 1~$\mu$m, respectively. White dashed rectangle represents the boundaries of ASMS's unit cell. White arrow defines the direction that corresponds to polarization angle $\theta$ = 90~$^{\circ}$. {\bf c,} Detailed photovoltage $V_{\rm ph}$ investigation on radiation polarization angle $\theta$ and gate voltage $V_{\rm g}$. Zoomed curve at $\theta$~=~0$^\circ$ is shown on inset. {\bf d,} Absolute value of polarization ratio dependence on $V_{\rm g}$ showing how much the signal with polarization along the wedges exceeds the signals with other polarizations. The value at the maximum responsivity point $V_{\rm g}$~=~41.5~V is indicated. Inset shows the resistance $R$ dependence on $V_{\rm g}$. {\bf e-j,} Simulations of intensity enhancement ({\bf e,f}), carrier temperature ({\bf g,h}) and generated photocurrents ({\bf i,j}) for wedge-type unit cell of our ASMS device in two orthogonal polarizations of incident radiation $\theta$ = 0~$^{\circ}$ and $\theta$ = 90~$^{\circ}$. Color bars on \textbf{e} and \textbf{f} have a linear scale up to 1 and a power scales starting from 1. On photocurrent density streamplots \textbf{i} and \textbf{j} the total photocurrent across $x$ direction through the right boundary of unit cell $I_{\rm ph}^{x}$ are indicated.}
    \label{fig3}
\end{figure*}

We confirm the additive character of photovoltages generated by individual stages by comparing the response of the multi-stage devices in Fig.~\ref{fig2}c. The signal of $w_9$-device is considerably larger than that of $w_5$, while the signal of $w_{11}$ is comparable to that of $w_9$. This is explained by capturing larger amount of optical power by larger detectors, and saturation of power for devices larger than the beam area.

The role of keen wedge in the formation of photoresponse becomes clear upon inspection of the polarization-dependent photovoltage signal. Fig. \ref{fig2}b shows the dependencies of $V_{\rm ph}$ on polarization angle $\theta$ (counted off from the vertical linear metal contact) at various gate voltages $V_{\rm g}$ (similar dependencies for $w_5$ and $w_{11}$ see in Fig.~\ref{figS2}). The photovoltage possesses a sharp maximum at $\theta = 90^\circ$, i.e. at electric field vector along the wedge direction. Depending on carrier density, the polarization ratio (PR) varies in the range $2...3$, and the signal is always maximized for the polarization along the wedge. Such dependence is naturally explained by the quasi-electrostatic lightning-rod effect.

{\bf Optimizing the responsivity and polarization ratio.} We suggest that sharper wedge angle results in stronger field concentration~\cite{LL-electrodynamics}, higher photovoltage, and higher PR. 
In the same line, it's tempting to minimize the size of individual cells comprising an ASMS. Indeed, the photovoltage in graphene-metal devices is generated only in a narrow Schottky junction layer $l_{\rm J} \sim 50$ nm~\cite{Near-field-microscopy-photocurrent,Tielrooij_Femtosecond_photoresponse}, and all the remaining graphene area is redundant. To harvest photoelectricity efficiently, it is tempting to make the source-drain distance comparable to the junction length.

The layout of empirically optimized device $w$ is shown in Fig.~\ref{fig3}a,b. Its unit cell has length $L=2.5$ $\mu$m and width $W=0.4$ $\mu$m, the wedge height is $h=930$ nm and full wedge angle is $\alpha \approx 15^\circ$. The inter-electrode gap is 160~nm and comparable with the Schottky junction length. The full ASMS of area 30 $\mu$m $\times$ 30 $\mu$m includes 11 cells in series and 75 cells in parallel. Scanning electron microscopy (Fig.~\ref{fig3}b) confirms persistent wedge sharpness with curvature radius $r_{\rm c} \lesssim 50$ nm.

The results of photovoltage measurements shown in Fig.~\ref{fig3}c demonstrate that shrinkage of ASMS cell and sharpening of the wedge indeed result in larger photoreponse. The photovoltage reaches 340 $\mu$V at $P_0~=~3.04$~mW laser power, which in three times exceeds the responsivity of non-optimized device despite three times smaller area (see details in Table~\ref{resp_table}). Another important observation is very high polarization ratio of the optimized device order of 200 (Fig.~\ref{fig3}d)~\footnote{More precisely, PR equals $-200$ as the photovoltage changes sign at $\theta \sim 5^\circ$. However, the sign is maintained at most polarization angles.}. High PR is maintained in a stably broad range of carrier densities, while at some selected densities it tends to divergence~\cite{Semkin_APL,Wei_Configurable}. Such high polarization ratios have not been achieved in graphene-metal~\cite{wedge_zhang_2024} or graphene $p-n$ junction detectors before~\cite{Doping_PNJ}, which underpins the importance of lightning-rod effect at the wedge-shaped contact.

Estimating the ASMS photodetectors' responsivity, we evaluate the infrared power reaching the device with area of $30 \times 30$ $\mu$m$^{2}$ as $P_{0} = 880$~$\mu$W (see Methods). For optimal gate voltage $V_{\rm g} = +41.5$~V in Fig.~\ref{fig3}c, the voltage responsivity equals $r_{\rm V} = 383$ mV/W. As all devices operate in the zero-bias mode, their noise spectral density $s_v$ is pretty low and governed by Johnson-Nyquist noise with $s_v = (4 kT R)^{1/2} \approx 19$ nV/Hz$^{1/2}$. The resulting noise equivalent power equals 50 nW/Hz$^{1/2}$ and is comparable to that of contemporary uncooled infrared detectors. 

It is instructive that the voltage responsivity can be dramatically magnified if all elementary cells of the metasurface detector are connected in series. In such a structure, the responsivity enhancement factor equals the number of elementary cells in a column $N_{\rm c} = 75$, thus the maximum achievable responsivity $r^{\max}_{\rm V} = 28.7$ V/W. A similar enhancement of current responsivity can be achieved for parallel arrangement of elementary cells. 

{\bf Detection physics and design rules.} The shape of carrier density-dependent photovoltage curves, Fig.~\ref{fig3}c, resembles much the density dependence of Seebeck coefficient of graphene. This fact, along with previous studies~\cite{Hot-carriers-review,Castilla_Asymmetric_Heating}, points to the central role of photo - thermoelectric effect in the device photoresponse. More precisely, enhancement of local electromagnetic intensity near the wedge apex heats up the adjacent metal-graphene Schottky junction. The junction at the linear metal-graphene contact is subjected to weaker field, and remains 'cold'. The heating imbalance at the two junctions drives the total thermoelectric photocurrent.

To establish further optimization strategies, we have developed a multi-stage model for thermoelectric detector (see Methods for details). As a first step, it includes the simulation of electromagnetic (Joule) absorption using the commercial Microwave Studio package. The optical conductivity of graphene is assumed to take its 'universal' value $\sigma_{\rm opt} = e^2/4\hbar$. In the next step, the hot carrier temperature is obtained by solving the heat diffusion problem with Joule-type source. Finally, we evaluate the photo-thermoelectric current by solving the 2d Poisson equation for the photoinduced electric potential~\cite{Levitov_SR_for_photocurrent}.

The simulated profiles of local fields, carrier heating, and photocurrent are shown in Fig.~\ref{fig3}e-j for the two orthogonal radiation polarizations. The ${\bf E}$-field for polarization along the wedge is greatly enhanced at its apex, with enhancement factor $\sim 500$ for the local intensity. The asymmetry of absorbance translates in dissimilar electron temperatures at the left and right Schottky junctions. Interestingly, the junction temperature difference at the symmetry axis of device is relatively small ($\delta T \approx $~8.3~K). Still, it drives a global electron flow from the hot wedge to the colder linear contact. A part of the flow is also directed to the wedge base that persists at the lower carrier temperature. Partial graphene etching can suppress this backflow~\cite{Wei_Spin_light} and thus enhance the photoresponse.

As the incident ${\bf E}$-field turns orthogonal to the wedge, the field enhancement turns to the field suppression. The resulting raise in electron temperature is quite small, $\delta T^{\parallel}_{\rm max} = 0.6$ K for this polarization against $\delta T^{\perp}_{\rm max} = 27$~K for the optimal one. This results in a very weak photocurrent.

Further numerical inspection of simulation results shows that neither absorbance variations nor temperature variations with light polarization explain the high polarization ratio ${\rm PR} \approx 200$. Indeed, the absorbance of wedge-shaped structure in the two crossed polarizations equals $3.9$ \% and $0.1$ \%, as computed from field-enhancement maps in Fig.~\ref{fig3}e,f. The ratio of absorbances is approximately five times below the polarization ratio. The ratio of maximum carrier temperatures $\delta T^{\parallel}_{\rm max}/\delta T^{\perp}_{\rm max} = 45$ almost follows the ratio of absorbances. High PRs are reproduced only at the stage of the drift-diffusive thermoelectric modeling. 

\begin{figure}[ht!]
    \includegraphics{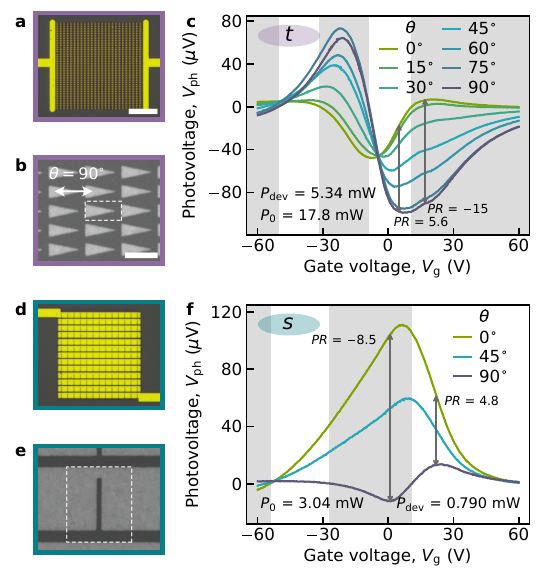}
    \caption{{\bf Configurable photodetectors based on alternative asymmetric metasurfaces.}  {\bf a,b,d,e,} Images of graphene detectors with triangle-type metasurface ({\bf a, b}) and with slit-type metasurface ({\bf d, e}). The scale bar in optical images {\bf a} and {\bf c} equals 10 $\mu$m, in SEM images {\bf b} and {\bf d} it equals 1 $\mu$m. {\bf c,f,} The density and polarization dependencies of the triangle-type device ({\bf c}) and the slit-type device ({\bf f}) photoresponse. White and gray areas represent ranges of positive and negative polarization ratios, respectively.}
    \label{fig4}
\end{figure}

The above observations imply that the photocurrent is sensitive not only to the magnitude of heating $\delta T$, but also to the position of the hot spot. For efficient photocurrent collection, the hot electrons should be generated at the points where they're easy to extract. This statement can be considered as a design rule for photo-thermoelectric detectors. In case of the parallel polarization, electrons get hot at the wedge-shaped source and are readily collected by the adjacent cold drain. In the case of orthogonal polarization, hot electrons are locked in the narrow inter-wedge slits, and hardly reach the drain. The effect is analogous to the increase in saturated vapor pressure for droplets with large positive curvature~\cite{Saturated_Vapor_Kelvin}. A direct proof of strong photocurrent sampling near the wedges is obtained by solving the 2d Poisson equation for the photoinduced electric potential (see Methods, eq.~(\ref{eq-solution})).

{\bf ASMS detectors with configurable polarity.} Two extra asymmetric metasurface detectors with a different geometry support the derived design rules. The device labeled as $t$ (triangle) is shown in Fig.~\ref{fig4}a,b. It possesses a metasurface of unconnected metallic triangles above the graphene. The device labeled as $s$ (slit) is shown in Fig.~\ref{fig4}d,e; its metasurface represents a series of metal stripes with asymmetrically cut slits. The absorbance for $s$-device reaches $2.8$ \% for polarization parallel to the slit, and $1.1$ \% for polarization orthogonal to the slit. For $t$-device, it reaches $3.8$ \% for polarization along the triangle axis and $1.7$ \% for the orthogonal direction.

Both devices, though having high absorbance and built-in asymmetry, demonstrate smaller photoresponse as compared to the optimized ASMS of Fig.~\ref{fig3}. These results are confirmed in Fig.~\ref{fig4}c,f. The peak photoresponses $s$- and $t$-type devices are three and twenty times smaller, as compared to the wedge-type device. The reason can be understood from hot carrier flow profiles. In the $s$-structure, hot electrons are locked in the slit and hardly reach the linear contact. In the $t$-structure, hot electrons at the triangle apex are spatially frustrated. They can either flow to the subsequent cell, or counter-flow to the triangle base.

Despite moderate photoresponse, the alternative devices have an intriguing property of gate-controlled polarization selectivity. More precisely, they can discriminate between two orthogonal light polarizations via different sign of photovoltage at specific $V_{\rm g}$ values (gray-shadowed regions in Figs.~\ref{fig4}c,f). By changing the gate voltage, one achieves polarization-independent sign of the photoresponse. Finer tuning of gate voltage enables to select between positive ${\rm PR}$ either close to unity or well above unity (Fig.~\ref{figS3}).

The gate tuning of ${\rm PR}$ can be related to the abrupt variations of local intensity in detectors with sharp wedges. Indeed, no such gate-configurable polarity was observed in graphene detectors with triangular antidot pattern on a gate~\cite{Ganichev_Patterned_gate}. With a similar lattice symmetry to our $t$-detector, their local fields were not enhanced as the pattern was spatially displaced from the channel. We can suggest that gate tuning of PR stems from the change in Schottky junction length with varying the gate voltage. Changing the $V_{\rm g}$, one scans the light-sensing junction along the highly-non uniform local intensity profile. In this scenario, the sign of photovoltage depends not only on the carrier polarity in graphene, but also on the spatial location of the junction.

{\bf Conclusion.} Our experimental findings complete the design rules for two-dimensional thermoelectric detectors, which previously relied on high absorbance and built-in asymmetry. A previously overlooked requirement is generation of hot carriers at the convexities (singularities) of metal-2d interface. This facilitates the hot carrier collection by the cold contacts and finally enhances the photocurrent. Factors limiting the local field enhancement in singular metasurfaces include finite curvature of wedges, finite conductivity of underlying 2d system~\cite{Nikulin2021_edge_diffraction}, and non-locality of conductivity in metal and graphene~\cite{Pendry_Limits_plasmonics}.

While the physical model for gate-tunable PRs in ASMS detectors is yet to be developed, the practicality of the designed devices is already at high level. Both positive and  negative PRs are feasible near the maxima of photoresponse, as compared to prior observations at nearly zero photovoltage close to charge neutrality~\cite{Semkin_APL,Wei_Configurable}. Our detectors are largely insensitive to material quality due to the quasi-electrostatic nature of the lightning-rod enhancement. This contrasts to the plasmonic~\cite{Castilla_Asymmetric_Heating,Bandurin_resonant,Muravev_plasmonic_detector} and antenna-type~\cite{Koepfli_Metamaterial,Wei_Calibration_free,Wei_Spin_light} field enhancements. We anticipate some residual spectral dependence of ASMS detectors' responsvity stemming from wavelength-dependent length of field-enhancement region $l_{\rm enh} \sim (0.01...0.1)\lambda_0$. From this perspective, the ASMS detectors are best suited for mid- and far-infrared light, where the domain of field enhancement largely overlaps with the metal-2d Schottky junction. 

\subsection*{Methods}
\textbf{Responsivity estimates.}
The power in the device plane was measured with a Thorlabs S302C thermal power head connected to a PM100D powermeter. We replaced the devices holder with the thermal head to preserve the optical path. The measured average laser power (50 \% duty cycle) varies in the range $P_0^{\rm avr} = 1.52 \dots 8.90~{\rm mW}$, depending on the laser current, while the power amplitude is    $P_0 = 3.04 \dots 17.8~{\rm mW}$. The power per device area  with dimensions $L$ and $W$, illuminated by a Gaussian focused beam $p(x) \propto e^{-(x^2+y^2)/2\sigma^2}$ is
\begin{equation}
	P_{\rm dev} = \operatorname{erf}\left(\frac{L}{2\sqrt{2}\sigma}\right)\operatorname{erf}\left(\frac{W}{2\sqrt{2}\sigma}\right) P_0.
\end{equation}
Our laser spot has $\sigma$ = 20 $\mu$m, thus the $w$-, $t$- and $s$-devices with $L=W=30$ $\mu$m receive 30 \% of total laser power.

The amplitude of photovoltage under steady illumination $V_{\rm ph}$ is calculated from lock-in readings $V_{\rm LI}$ under meander-modulated illumination using 
\begin{equation}
    V_{\rm ph} = \frac{\pi}{\sqrt{2}}\,V_{\rm LI},   
\end{equation}
The responsivities per total power ($r_{\rm V}$) and per power reaching the device ($r_{\rm V,\,dev}$) are
\begin{equation}
    r_{\rm V} = \frac{V_{\rm ph}}{P_0} \qquad {\rm and} \qquad r_{\rm V,\,dev} = \frac{V_{\rm ph}}{P_{\rm dev}}.
\end{equation}

\textbf{Modeling.} Electromagnetic modeling was implemented in Microwave Studio package with single-frequency finite element solver. The dielectric functions of contact metals and surrounding dielectrics were taken from refractiveindex.info database. 

Modeling of thermal transport is based on heat conduction equation for electron temperature $T({\bf r})$:
\begin{equation}
\label{eq-thermal}
    -(\nabla, \chi_e({\bf r}) \nabla T({\bf r})) = -\frac{C_{e}}{\tau_\varepsilon} [ T ({\bf r}) - T_0] + \frac{1}{2}\sigma_{\rm opt} |{\bf E}|^2,
\end{equation}
here $\chi_e$ and $C_{e}$ are the electronic thermal conductivity and heat capacity, respectively, $\tau_\varepsilon$ is the energy relaxation time due to substrate phonons, $T_0$ is equilibrium temperature. The last term in Eq. (\ref{eq-thermal}) is the Joule heating by electromagnetic field.

Modeling of photo-thermoelectric response is based on continuity relation supplemented with microscopic expression for electron current. The latter includes thermal diffusion and drift components. The resulting equation on light-induced electric potential $V({\bf r})$ is (see Supplementary Information, Section II for derivation):
\begin{equation}
\label{eq-thermoelectric}
    (\nabla, \alpha_e({\bf r}) \nabla T({\bf r}) - \sigma_e ({\bf r}) \nabla V({\bf r})) = 0,
\end{equation}
here $\alpha_e({\bf r})$ is the thermal diffusivity and $\sigma_e({\bf r})$ is the electron conductivity. The three functions $\alpha_e$, $\chi_e$ and $\sigma_e$ are bound by the Wiedemann-Frantz and Mott relations. Only one of them retains free, which we choose to be $\sigma_e = \mu e (n+p)$.

Equation (\ref{eq-thermoelectric}) can be solved semi-analytically for the photocurrent $I_{\rm ph}$ under assumption of uniform conductivity $\sigma_e ({\bf r}) = {\rm const}$. The result reads as
\begin{equation}
\label{eq-solution}
    I_{\rm ph} = \frac{1}{V_0} \int{d{\bf r} ({\bf E}_0({\bf r}), \alpha_e({\bf r}) \nabla T({\bf r}))},
\end{equation}
where ${\bf E}_0$ is dc the electric field in the channel upon application of potential $V_0$ to the drain (not to be confused with local radiation field ${\bf E}({\bf r})$). Eq.~({\ref{eq-solution}}) shows that photocurrent sampling is most efficient from the points with strong local dc electric field, i.e. near the edges.

\subsection*{Acknowledgements}
Device fabrication was supported by the grant \# 020-11-20211446 ''Development of graphene nanomaterial fabrication technology for applications in optoelectronic components'' on assignment of JSC ''Skanda Rus''. Optoelectronic measurements and modelling was supported by the grant \# 24-79-10081 of the Russian Science Foundation. The devices were fabricated using the equipment of the Center of Shared Research Facilities (Moscow Institute of Physics and Technology).

\subsection*{References}
\bibliography{references}

\onecolumngrid
\renewcommand{\theequation} {S\arabic{equation}}
\renewcommand{\thefigure} {S\arabic{figure}}
\renewcommand{\thetable} {S\arabic{table}}
\setcounter{figure}{0}
\newpage

\section*{Supporting Information}
\section{Extra device parameters and results of photovoltage measurements}

\begin{figure*}[ht]
    \includegraphics[width=0.95\textwidth]{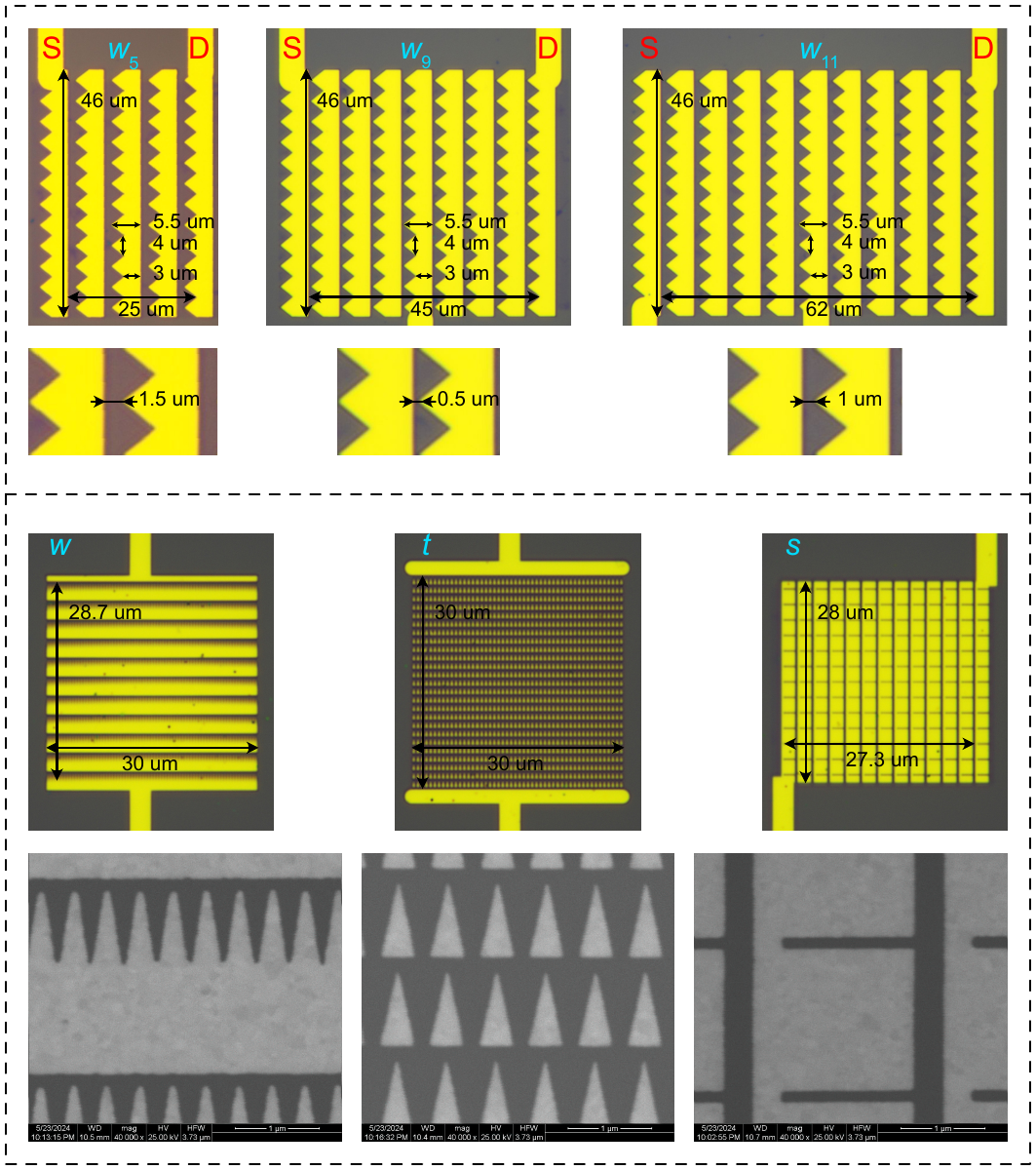}
    \caption{\label{figS1}{\bf Geometrical parameters of devices.} Optical and scanning electron microscopy images of all investigated devices: $w_5$, $w_9$, $w_{11}$, $w$, $t$, $s$.}
\end{figure*}

\begin{figure*}[ht]
    \includegraphics[width=\textwidth]{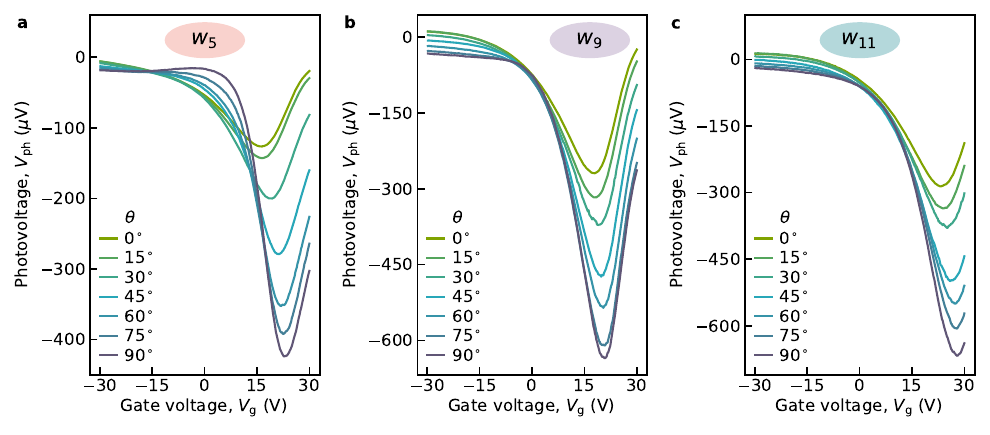}
    \caption{\label{figS2}{\bf Photoresponce.} Gate- and polarization dependencies of the photovoltage for $w_5$ ({\bf a}), $w_9$ ({\bf b}) and $w_{11}$ ({\bf c}) devices.}
\end{figure*}

\begin{figure*}[ht]
    \includegraphics[width=\textwidth]{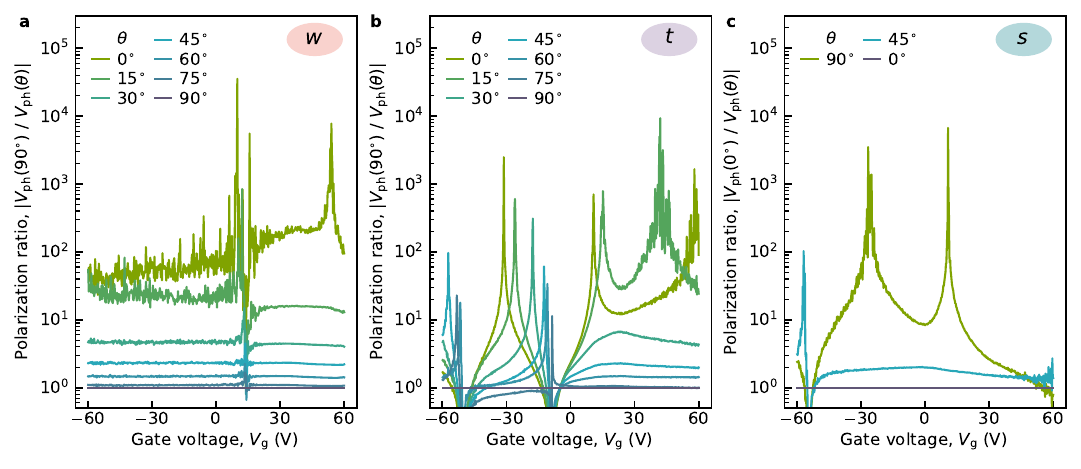}
    \caption{\label{figS3}{\bf Polarization ratio.} Absolute value of polarization ratio dependencies on gate voltage for $w$ ({\bf a}), $t$ ({\bf b}) and $s$ ({\bf c}) devices.}
\end{figure*}

\clearpage
\begin{table*}[ht]
    \includegraphics[width=\textwidth]{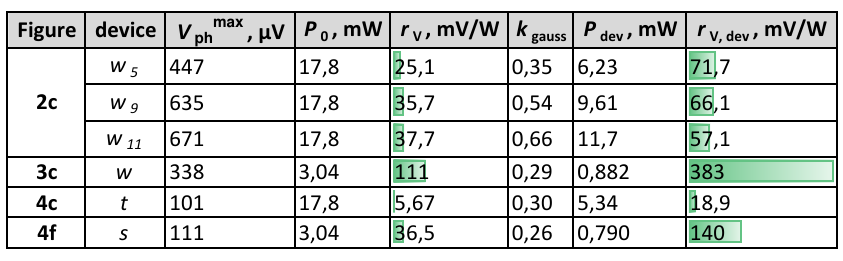}
    \caption{{\bf Responsivity.} Responsivities of all discussed devices, estimated at the total laser spot power ($r_{\rm V}$) and at the power per device area ($r_{\rm V,~dev}$).}
    \label{resp_table}
\end{table*}

\section{Solution of Poisson equation for the photoinduced potential}

An equation governing the distribution of photoinduced electric potential is obtained from current continuity relation
\begin{equation}
\label{eq-thermoelectric1}
    (\nabla,\mathbf{j({\bf r})}) = 0,
\end{equation}
supplemented with microscopic experssion for the current density
\begin{equation}
   \mathbf{j}({\bf r})=\alpha_e({\bf r}) \nabla T({\bf r}) - \sigma_e ({\bf r}) \nabla V({\bf r}). 
\end{equation}
here $\alpha_e({\bf r})$ is the thermal diffusivity and $\sigma_e({\bf r})$ is the electron conductivity. We assume $\sigma_e({\bf r}) = {\rm const}$. Merged together, the two above equation yield
\begin{equation}
\label{eq-thermoelectric2}
    (\nabla, \alpha_e({\bf r}) \nabla T({\bf r}) - \sigma_e ({\bf r}) \nabla V({\bf r})) = 0,
\end{equation}

For illustrative purposes, we assume uniform channel conductivity $\sigma({\bf r})={\rm const}$. Its solution can be obtained with the technique of characteristic potentials. The characteristic potential $v({\bf r})$ is formed upon application of potential $V_0$ to the drain, and grounding of the source. After taking into account ~eq.(\ref{eq-thermoelectric1}) and boundary conditions for $v({\bf r})$, integration of identity
\begin{equation}
\label{eq-vj}
    \nabla(v({\bf r}) \mathbf j({\bf r}))=(\nabla v({\bf r}),\mathbf{j}({\bf r}))+v({\bf r})(\nabla,\mathbf{j}({\bf r}))
\end{equation}
over the domain and applying divergency theorem yields 
\begin{equation*}
\label{eq-vjInt}
    I_{ph}V_0= \int{d{\bf r} ({\bf E}_0({\bf r}), \mathbf{j}({\bf r}))}=\int{d{\bf r} ({\bf E}_0({\bf r}), \alpha_e({\bf r}) \nabla T({\bf r}))}-\int{d{\bf r} ( {\bf E}_0({\bf r}),\sigma_e({\bf r}) \nabla V({\bf r}))}.
\end{equation*}
Due to the uniformity of conductivity, the second  integral in is zero. Indeed, 
\begin{equation*}
\label{eq-zero}
-\int{d{\bf r} ( {\bf E}_0({\bf r}), \nabla V({\bf r}))}=\int{d{\bf r} ( \nabla v({\bf r}), \nabla V({\bf r}))}=    \int{d{\bf r} ( \nabla, V({\bf r}) \nabla v({\bf r}))}=0.
\end{equation*}
Here we use the identity $$( \nabla, V({\bf r}) \nabla v({\bf r}))=( \nabla v({\bf r}), \nabla V({\bf r}))+V({\bf r})\Delta v({\bf r}),$$ the divergence theorem and the boundary conditions for $V({\bf r})$.

This justifies Eq.~(3) of the main text. Even simpler form can be obtained for a stepwise modulation of thermal  diffusivity. To this end, we integrate the identity 
\begin{equation}
\label{eq-TE0}
    \nabla(T({\bf r}) \mathbf E_0({\bf r}))=(\nabla T({\bf r}),\mathbf{E}_0({\bf r}))+T({\bf r})(\nabla,\mathbf{E}_0({\bf r}))
\end{equation}
over the domain and divergence theorem gives 
\begin{equation}
\label{eq-E0T}
\int{d{l}  (\mathbf{E}_0({\bf r}),\mathbf{n})T({\bf r})}=\int{d{\bf r} ({\bf E}_0({\bf r}),\nabla T({\bf r}))}.
\end{equation}

Here the first integral is taking over the boundary of domain, $\mathbf{n}$ is an outward normal for the domain.
 Taking into account discontinuity of $\alpha_e({\bf r})$, we can apply eq.~(\ref{eq-solution}) with eq. (\ref{eq-E0T}) to different subdomains, on each of them $\alpha_e({\bf r})=\rm const$, and then simply add these integrals. Applying this procedure we get
\begin{equation}
\label{eq-solution1}
    I_{\rm ph} = \frac{1}{V_0} \sum_i\alpha_{e i}\int{d{l}  (\mathbf{E}_0({\bf r}),\mathbf{n})T({\bf r})},
\end{equation}
 where integral is taken over the boundary of each subdomain.

\end{document}